\documentclass[pdflatex,sn-mathphys-num]{sn-jnl}

\usepackage{graphicx}%
\usepackage{multirow}%
\usepackage{amsmath,amssymb,amsfonts}%
\usepackage{amsthm}%
\usepackage{mathrsfs}%
\usepackage[title]{appendix}%
\usepackage{xcolor}%
\usepackage{textcomp}%
\usepackage{manyfoot}%
\usepackage{booktabs}%
\usepackage{algorithm}%
\usepackage{algorithmicx}%
\usepackage{algpseudocode}%
\usepackage{listings}%

\usepackage[nolist]{acronym} 
\usepackage{todonotes}
\setuptodonotes{inline}
\usepackage{textgreek}

\usepackage[T1]{fontenc}

\usepackage{hyperref}

\theoremstyle{thmstyleone}%
\theoremstyle{thmstyletwo}%

\theoremstyle{thmstylethree}%

\begin{document}

\title[Article Title]{Quality-Diversity Search in Sound Generation: Investigating Innovation Engines for Audio Exploration}


\author*[1]{\fnm{Björn Þór} \sur{Jónsson}}\email{bthj@uio.no}

\author[1]{\fnm{Çağrı} \sur{Erdem}}\email{cagrie@uio.no}
\equalcont{These authors contributed equally to this work.}

\author[2]{\fnm{Stefano} \sur{Fasciani}}\email{stefanof@uio.no}
\equalcont{These authors contributed equally to this work.}

\author[1]{\fnm{Kyrre} \sur{Glette}}\email{kyrrehg@uio.no}
\equalcont{These authors contributed equally to this work.}

\affil*[1]{\orgdiv{Department of Informatics}, \orgname{University of Oslo}, \orgaddress{\country{Norway}}}

\affil[2]{\orgdiv{Department of Musicology}, \orgname{University of Oslo}, \orgaddress{\country{Norway}}}

\abstract{
This study addresses the challenges composers and sound designers face in creating and refining tools to achieve their musical goals. Using evolutionary processes to promote diversity and foster serendipitous discoveries, we automate the search through uncharted sonic spaces for sound discovery, arguing that diversity-promoting algorithms can bridge the gap between the theoretical realisation and practical accessibility of sounds. We describe a system for generative sound synthesis combining Quality Diversity (QD) algorithms with a supervised discriminative model, inspired by the Innovation Engine algorithm, and explore different configurations and the interplay between the chosen synthesis approach and the discriminative model. We examine the interaction between Compositional Pattern Producing Networks (CPPNs) and Digital Signal Processing (DSP) graphs, introducing a novel approach that uses multiple specialised CPPNs for different frequency ranges; this yields simpler networks while maintaining performance comparable to single-CPPN setups. We also investigate evolutionary stepping stones by analysing goal switches between musical and non-musical contexts, revealing how lineages traverse unlikely paths to current elites. Expanding the behaviour space of a previous study to include various sound durations, we uncover specialisation within temporal niches. Results indicate that CPPN and DSP graphs coupled with a Multi-dimensional Archive of Phenotypic Elites (MAP-Elites) and a deep learning classifier can generate a substantial variety of synthetic sounds, diverse and innovative across temporal and contextual dimensions. We present the generated sound objects through an online explorer and as rendered sound files, and, in the context of music composition, an experimental application that showcases their creative potential across various durations and contexts.
}

\keywords{Sound Synthesis, Quality Diversity Search, Innovation Engines}



\maketitle

\acresetall

\section{Introduction}

Either you know what sound you’re looking for, or you don’t know what sound you’re looking for. In the latter case, inquiry, or prompting, is impossible. To discover new sounds, you must recognize them when you have found them. But if you can do that, you must have known them already. Transferring such a paraphrasing \cite{noe_entanglement_2023} of Meno’s Paradox to the domain of novel sound design can be a way of establishing the usefulness of serendipitous sonic discoveries, where a new sound may not have been explicitly sought after but immediately recognised when heard.    
With all sound admissible as material for making music and all sounds theoretically possible with digital synthesis,
there is still much more to explore considering the entirety of the sonic domain 
\cite{wyse_free_2003}.
Composers and sound designers often need to create and refine new tools in order to achieve their musical goals. This endeavour may be hindered by a lack of technical expertise.
Our proposed approach leverages evolutionary processes to generate novel sounds, thereby facilitating the creative journey and overcoming the technical barriers that may limit composers and sound designers in expanding their sonic repertoire. 

We work towards an approach to automate 
navigation through 
previously unexplored sonic territories. As such, while entirely novel, the discovered sounds can be perceived as appealing and seemingly recognisable to the listener despite their unprecedented nature.
Such investigations have been carried out interactively in the visual domain \cite{secretan_picbreeder_2011}, demonstrating the usefulness of abandoning specific objectives, or at least 
switching goals as stepping stones are found while traversing paths to interesting discoveries. These findings provided a basis for proposing the Novelty Search algorithm \cite{lehman_abandoning_2011} and later other variants, forming a family of \ac{QD} search algorithms 
\cite{Lehman2011,mouret_illuminating_2015,pugh_quality_2016,cully_quality_2018}.
These \ac{QD} algorithms combine the open-endedness of Novelty Search with competition between solutions in their own “niche”, resulting in 
diverse and 
high-performing (quality) solutions. 
Overall, QD algorithms serve as effective tools for illuminating high-quality solutions within a domain and are powerful search algorithms in their own right. This is due in part to their ability to exploit behavioral diversity and stepping stones during the search process, which can lead to discovering a variety of valuable solutions \cite{gaier_are_2019,nordmoen-frontiers2021}.
%
%
To drive automated exploration with such diversity-promoting algorithms, the Innovation Engine algorithm abstracts the process of human curiosity, replacing human judgement with a discriminative model that identifies interesting ideas \cite{nguyen_innovation_2015,nguyen_understanding_2016}. 
Innovation Engines integrate two key components: \acp{EA}, such as those from the family of \ac{QD}, capable of generating and gathering various novel outputs; and a model capable of distinguishing that novelty and evaluating its quality, such as \acp{DNN}, creating niches and competition within them, thus providing selection pressure to quide \ac{QD} search.
The ultimate goal of this architecture is to continuously generate interesting and innovative creations in any given field.

\begin{figure}
\centering
\includegraphics[width=\textwidth]{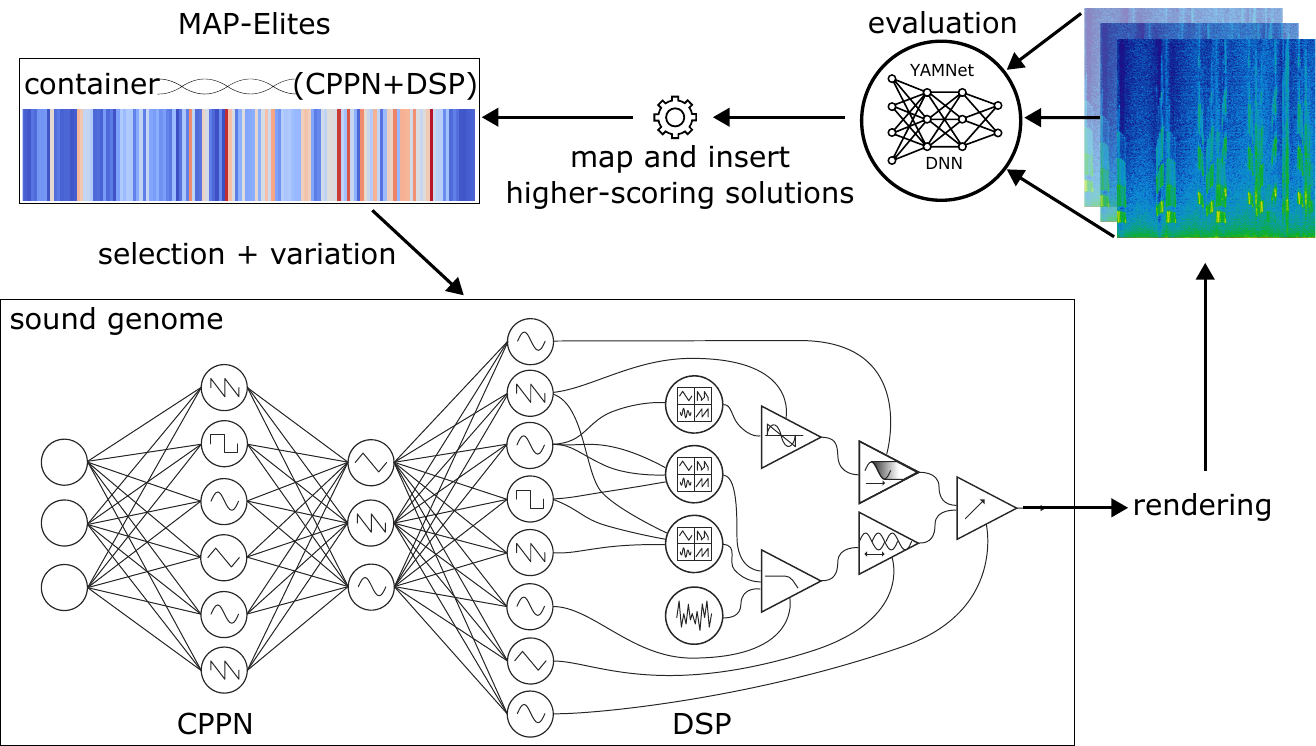}
\caption{The QD algorithm MAP-Elites uses 
the 
pre-trained 
YAMNet DNN classifier to define cells in a container and the performance of an evaluation candidate across those cells to determine placement and replacement in that archive. The genome of each evaluation candidate is rendered to a waveform, which is supplied to the classifier.
Inputs to the CPPN are discussed in section \ref{sec_approach_periodic_signal_composition}.
}
\label{fig:sound-innovation-engine-pipeline}
\end{figure}

\acp{CPPN} \cite{stanley_compositional_2007} are a foundation of the explorations leading to the Novelty Search and Innovation Engine algorithms. The networks abstract unfolding development in evolutionary processes, which build a phenotype over time.
This is done by using any variety of canonical functions at each node, based on the idea that the order in which the networks compose functions can 
provide that abstraction.
This can be compared with the process of timbral development, where musical expression depends on changes and nuances over time. 
The use of 
patterns produced by \acp{CPPN} as sources of sound- and control signals 
for sound synthesis has been explored in a novelty seeking \ac{IEC} \cite{takagi_interactive_2001} configuration, 
which was 
inspired by previous work 
on the generation of visual artefacts 
\cite{jonsson_interactively_2015}.
The 
representation of 
temporal unfolding provided by
\acp{CPPN} has been combined with the evolution of \ac{DSP} graphs during several iterations of investigation, 
detailed in \cite{jonsson_system_2024}.
This resulted in a distinct approach to sound synthesis, where any combination of the two graphs, 
depicted in figure \ref{fig:sound-innovation-engine-pipeline},
can be rendered at any duration, revealing the sub-patterns encoded by \acp{CPPN} over varying periods of time.

Given the more diverse application of sonic artefacts as material in creative processes, we argue that
there may be an even higher incentive to investigate the Innovation Engine algorithm's applicability in the sound domain. 
Furthermore, whereas humans can evaluate images in a split second, evaluation of sounds requires more time. 
There is a minimal duration threshold for perceiving salient features of sonic objects \cite{moore_hearing_1995} 
and we typically perceive them holistically as meaningful units in the 0.5 to 5 seconds range \cite{godoy_chunking_2009}.
Experiments with interactive evolution of sounds 
\cite{jonsson_interactively_2015} 
revealed how fatigue can set quickly in when potentially listening to a long series of 
taxing
sounds. 
This further limits the ability of humans to provide sufficient quantity of selection to have a significant effect on evolution.

Automating the discovery of new sounds is the goal of this study. We achieve this by applying the Innovation Engine algorithm to the sound synthesis approach developed in previous research on interactive novelty discovery.
By using the proposed technique for sound synthesis, the system does not need to be trained beforehand
as the evolutionary method starts from networks with no hidden nodes and progressively evolves 
primitive individuals by adding nodes and connections with the \ac{NEAT} algorithm \cite{stanley_evolving_2002}. 
In our initial experiments, we use a 
confidence
signal from a pre-trained discriminative model to guide \ac{QD} search, without human feedback in the evolutionary loop.
Investigating this setup
is intended to pave the way for further explorations of unbounded
discovery of interesting sounds.


Our contributions include researching the application of a special type of Innovation Engine in the sound domain with a distinct approach to sound synthesis within an \ac{EA},
extending our initial efforts towards understanding  sonic \ac{QD} landscapes previously published in \cite{johnson_towards_2024}.
Furthermore, we examine different configurations of our generative system and study how our sound synthesis method interacts with the discriminative model. 

In this extended study, we delve into the interaction between \ac{CPPN} and \ac{DSP} graphs in section \ref{sec:interplay_between_cppn_and_dsp_graphs}, presenting a novel approach that utilizes multiple specialised \acp{CPPN} for different frequency ranges. This configuration simplifies the \ac{CPPN} networks while maintaining performance on par with single-\ac{CPPN} setups. Furthermore, in section \ref{sec:context_switching} we explore evolutionary stepping stones by examining goal switches between musical and non-musical contexts, revealing how lineages navigate unlikely paths to reach their current elite states. Additionally, in section \ref{sec:behaviour_space_expansion} we investigate the temporal dimension of sound generation by extending the behavior space from the previous study in \cite{johnson_towards_2024} to encompass various sound durations, thereby uncovering specialization within temporal niches.

We also offer a web-based interface to explore the outcomes of our evolutionary processes through our Innovation Engine setup. 
Lastly, we showcase audio artefacts rendered from the solutions discovered during the \ac{QD} runs.

Experimental results, in the form of historical data from evolution runs, elite maps and genomes from each point in time, and sounds rendered from those genomes at final iterations, along with the source code to replicate the results, are available in the datasets accompanying this article \cite{jonsson_supporting_2024,jonsson_extended_2024}.






\section{Approach 
and Experimental Setup}
\label{sec_approach}






To start evaluating the applicability of the Innovation Engine algorithm in the domain of sounds, we combine 
a sound synthesis technique 
with a supervised discriminative model. 
%
The foundation of our sound-generating system relies on using 
the patterned 
outputs from \acp{CPPN} as the raw materials for sound and control signals. These signals can be utilised in their original form or further shaped through a \ac{DSP} graph. 
Such a design choice enables the 
evolutionary process to start from \textit{simple a beginning},
established with random initialization of the CPPN and DSP graph counterparts. This avoids dataset constraints that might limit the potential for discovery of novel sounds.
The genome evolved by the evolutionary (\ac{QD}) processes is composed of the \ac{CPPN} and \ac{DSP} networks and the evolvable connections between them. Details of this genome configuration are discussed and diagrammed in \cite{jonsson_system_2024}. Figure \ref{fig:sound-innovation-engine-pipeline} illustrates the data flow of our experimental setup and shows how the genome fits within the data pipeline.

\subsection{Behavioural Descriptor}
To guide the \ac{QD} search, we chose the \ac{YAMNet} \ac{DNN} classifier 
to define our search space.
The confidence scores output by the classifier for each class 
are used as selection signals for the \ac{QD} algorithm, as discussed in section \ref{sound_generation_and_qd_algorithm_variants}.
While this pre-trained network may limit our exploration, 
it was adopted in an effort to replicate a setup from 
previous evaluations of the Innovation Engine algorithm in the visual domain.
That classifier is trained on AudioSet \cite{gemmeke_audio_2017}, which can be 
considered as a sonic sibling of the \ac{DNN} classifiers trained on the ImageNet dataset \cite{deng_imagenet_2009}. \ac{YAMNet} outputs 521 scores from a logistic (softmax) layer, corresponding to AudioSet classes. 
The classifier's output is intended ``as a stand-alone audio event classifier that provides a reasonable baseline across a wide variety of audio events.''\footnote{
    YAMNet audio event classifier: \url{https://tfhub.dev/google/yamnet}
}. 
Our approach to sound generation can be somewhat likened to a unique type of sound synthesiser tjat is not crafted with the intention of mimicking natural sounds or creating textures that easily fit into well-known categories.
Many modern generative models excel at such tasks \cite{choi_foley_2023}, building on their prior training.
However, despite our synthesizer's limitations in rendering general sound events,
we considered the varied signal provided by 
YAMNet 
as a good starting point for driving the \ac{EA} towards diversity. We also considered it interesting to 
mirror the 
overall 
setup from experiments \cite{nguyen_innovation_2015,nguyen_understanding_2016,lehman_creative_2016} that inspire our sonic investigations.         


\subsection{Periodic Signal Composition}
\label{sec_approach_periodic_signal_composition}
One factor potentially influencing the search space is our choice of \ac{CPPN} activation functions and node types in the \ac{DSP} graph. \acp{CPPN} have commonly been used to compose Gaussian, sigmoid, and periodic functions,
such as in \cite{stanley_compositional_2007,secretan_picbreeder_2011}. In our case, the pattern-producing network can only compose periodic functions, commonly used as oscillators in a variety of sound synthesis techniques: sine, square, triangle, and sawtooth.
The node types in the \ac{DSP} graph are the same as in \cite{rice_gensynth_2015}, in addition to custom nodes, which 
were added in an effort to expand the sonic repertoire,
while at the same time 
widening 
the search space. 
Those additional nodes are a wavetable and a specialised additive synthesis node, where multiple audio signals are sourced from the \ac{CPPN} to fill a table in the former and represent partials or harmonics in the latter. The wavetable is traversed according to a control signal, also sourced from the \ac{CPPN}, in a manner similar to vector synthesis.
The partials in the additive synthesis node can be slightly inharmonic, according to a mutable parameter to each.

Although this augmentation of the sonic repertoire expands the search space, our model expresses some inductive bias through its structure by coupling sound sources from CPPN networks (in place of traditional oscillators) with DSP graphs. However, both parts of the structure evolve from scratch rather than being bootstrapped from established synthesizer circuits, which often requires our simulations to perform many evaluations to find high-quality sound objects.
Those limited constraints may lead to transformational discoveries.

The duration of sounds rendered from each genome is defined by a linear ramp of values from $-1$ to $1$ supplied to one \ac{CPPN} input, while the pitch is controlled by the rate of a periodic (sine) signal at another input. 
Velocity
is intended to simulate stimuli of different intensities when interacting with physical instruments, which is achieved by scaling the sine wave input by a velocity factor. The inputs are sampled at the same rate as the sampling rate of the audio graph.

\subsection{QD Algorithm}
For the diversity-promoting algorithm, we chose \ac{MAP-Elites} \cite{mouret_illuminating_2015}. Our experiments are based on a bespoke implementation of that algorithm, with the common addition of biasing it away from exploring niches that produce fewer innovations. 
This is achieved by assigning each niche a decrementing counter, representing a \textit{curiosity score} as defined in \cite{cully_quality_2018} with constants set as in \cite{lehman_creative_2016}. The counters start at a fixed value of 10, impacting the probability of that niche being selected for reproduction.
The classification outputs of the discriminative model define the cells of the behaviour space which the \ac{QD} algorithm explores, where the performance at each niche is determined by the confidence values for each class. During our main runs of \ac{QD} search, evaluations were performed in batches of 32. 


\subsection{Parameter Search}\label{sec_parameter_search}
Considering the temporal dynamics of sounds, and that the underlying pattern generator of our sound synthesis engine (\ac{CPPN}) encodes sub-patterns that 
reveal over time, we performed preliminary experiments classifying sounds rendered at a different duration for each evaluation. 
One of the configurations involved 112 evaluations of each sound genome, rendering it to sounds of 4 durations, 7 pitch variations and 4 amplitudes. 
To explore other parameters of the \ac{QD} search, such as mutation rates and their balance between the \ac{CPPN} and \ac{DSP} genome counterparts, as well as graph and node addition or deletion rates, we conducted a manual parameter search. Due to the computationally intensive nature of the task, these runs were based on a limited selection of parameter values.
A comprehensive collection of plots from those runs can be found in the datasets accompanying this paper \cite{jonsson_supporting_2024,jonsson_extended_2024}. 
We found that evaluating sounds with a duration of half a second frequently led to the emergence of successful sound variants. Therefore, we decided to use this specific duration for assessing sounds in the \ac{QD} runs of our primary experiments.
Runs with node- and connection addition rates of 10\% and corresponding deletion rates of 6\% 
resulted in the best performance during our parameter search. 
As such, we ran with that
as our baseline
configuration along with equal probability of mutating each genome counterpart.

\subsection{Genealogical History Preservation}
To keep track of the history of all changes during the evolutionary runs we considered implementing bespoke handling of the differences (diffing), but the idea came up to use Git \cite{pathak_introduction_2020} for version control. 
Beforehand, we realised that our usage pattern might not align with Git's intended purpose. However, knowing that it is battle-tested software, we were tempted to try out this idea, which we did for the experiments reported here.
In short, there were pros and cons. 
The software handled multiple commits per second efficiently, which can be seen as an unusual usage pattern. Moreover, the lookup of the evolutionary state from any point during the course the simulations was fast.
The default Automatic Garbage Collection (GC) did not interact effectively with our rapid commit pattern, so we disabled it and opted for separate GC processes intermittently or after the simulations completed, which resulted in significant deflation of the data volume.
Given our successful usage of Git in this scenario, we continue including it in the pipeline of our experiments \cite{jonsson_synth-iskromosynth_2024,jonsson_synth-iskromosynth-cli_2024,jonsson_synth-iskromosynth-evaluate_2024,jonsson_synth-iskromosynth-render_2024}, while also considering other Git implementation alternatives.

\section{Results}




For our main experiment, we ran 10 independent runs of the MAP-Elites algorithm, with the rates discussed in section \ref{sec_approach} and behaviour evaluated by \ac{YAMNet} on 0.5-second sounds. Each run lasted for 300 thousand iterations, with a batch or generation size of 32. At the start of each run, 50 seed iterations were performed, which differ from the rest of the iterations in that each 
individual is initialised from scratch rather than mutating a randomly selected elite occupying any of the cells. 




\begin{figure}
    \centering
    \begin{tabular}{@{}c|@{}c}
        \includegraphics[width=0.49\textwidth]{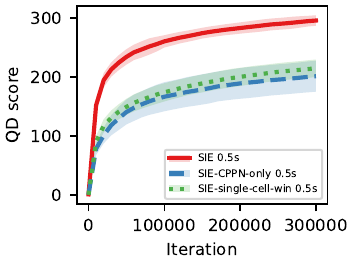}\label{qdScores} & 
        \includegraphics[width=0.49\textwidth]{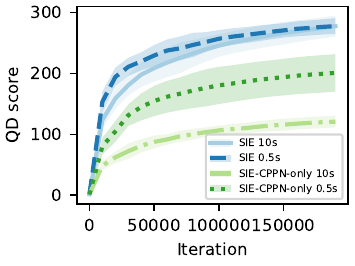}\label{qdScores_duration-comparison_basic-and-CPPNonly} 
    \end{tabular}
    \caption{
On the \textit{left}
scores are plotted for the baseline configuration (section \ref{sec_approach}), evaluating sounds of 0.5s duration, along with variants where sounds are only rendered from \ac{CPPN} mutations 
and where only one cell can be won at a time by each candidate elite. Data for each variant comes from 10 runs. 
The plot on the \textit{right}
compares the performance achieved when evaluating sounds rendered at two different durations---0.5s and 10s---from the baseline and \ac{CPPN}-only run configurations, each independently executed 5 times.
}
    \label{fig:qdScores}
\end{figure}

\subsection{Sound Generation- and QD Algorithm Variants}
\label{sound_generation_and_qd_algorithm_variants}
Aside from the set of evolution runs using our basic configuration described above, we performed two additional sets of runs. In one set, we altered the sound generation, and in the other set, we modified the progression of the \ac{QD} algorithm.



\subsubsection{Signal Processing Graph}
\label{sec:CPPN_plus_DSP_vs_CPPN_only}
To investigate the impact of merging \acp{CPPN} with \ac{DSP} graphs, we set up evolutionary runs in two distinct configurations: one in which an evolved \ac{CPPN} functioned solely as the audio signal source, providing a single output, and another where the \ac{CPPN} was paired with an evolving \ac{DSP} graph, allowing it to offer a multitude of audio and control signals, from up to 18 outputs. 

In our experiments, 
we quantify the QD algorithm performance by calculating 
the QD-score \cite{pugh_quality_2016,pugh_confronting_2015}.
This score is determined by summarising the confidence levels of the elites across the various classes delineated by YAMNet.
When comparing the results from these runs, 
with different definitions of behaviour by the sonic repertoire,
we observe in figures \ref{fig:qdScores} and \ref{fig:cellScores_dur_05} (section \ref{sec:Performance_Against_Pretrained_Reward_Signals}) that the phenotypes (i.e., sound objects) produced from the genomes where \acp{CPPN} and \ac{DSP} graphs were co-evolved achieved the highest overall QD-score, 
while the lower performance of the \ac{CPPN}-only runs is accompanied by higher complexity (plotted in figures figures \ref{fig:genomeStatistics_node_type_count} and \ref{fig:genomeStatistics} and discussed further in section \ref{sec:genome_complexity}).
Through informal listening sessions conducted by the authors, it was observed that the sounds rendered from runs where the evolution of \ac{DSP} graphs was allowed alongside \acp{CPPN} exhibited a higher degree of subjective aesthetic appeal. This phenomenon could potentially be attributed to the prevalence of classical synthesizer sounds, to which our ears have grown accustomed. In this context, the \ac{DSP} graph can be seen as functioning akin to a modular synthesizer patch, rendering us less inclined to perceive the raw output generated by \acp{CPPN} as inherently pleasing.
%
The rendered sounds can be auditioned 
in an online explorer (sec. \ref{sec:qualitative_evaluation}) or accompanying datasets \cite{jonsson_supporting_2024,jonsson_extended_2024}.


\subsubsection{Behaviour Space Coverage}\label{sec:multi_vs_single_cell_win}
The default behaviour of our \ac{MAP-Elites} implementation allows each evaluated individual to win all cells where it performs better or where there is a vacancy, so it reaches full coverage from the first seed. 
To examine the effect of gradually covering the map of cells by allowing each candidate to potentially win only one cell, the one where it receives the highest confidence from the classifier, we performed an identical set of runs except with that restriction in place.
Runs where at most one cell at a time is won reached a coverage of 57.4\%\textpm3.4\%, with their \ac{QD}-score following a trajectory similar to that of full coverage \ac{CPPN}-only runs, as depicted in figure 
\ref{fig:qdScores}, \textit{left}.

\subsubsection{Elite Populations}
\begin{figure}
    \centering
    \begin{tabular}{@{}c|@{}c}
        \includegraphics[width=0.49\textwidth]{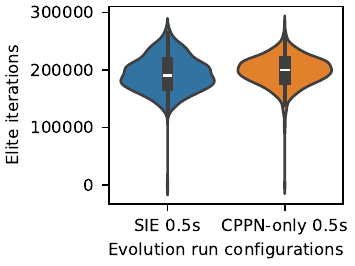}\label{fig:eliteGenerations} &
        \includegraphics[width=0.49\textwidth]{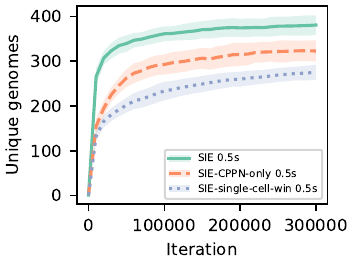}\label{fig:genomeSets}
    \end{tabular}
    \caption{
    \textit{Left}: Distribution of iteration numbers at which the current class elite was discovered. 
    \textit{Right}: Count of unique individuals, as it evolves through iterations of the evolution runs.
    }
    \label{fig:eliteGenerationsAndGenomeSets}
\end{figure}

Figure \ref{fig:eliteGenerationsAndGenomeSets} (\textit{left}) shows that the range of iterations where the current elites are found at the end of each run is sharply delimited around iterations 150K to 250K of the CPPN-only runs, while the \ac{CPPN}+\ac{DSP} runs continue to discover new elites more gradually throughout the latter half of the runs. 

The set of unique elites at the end of \ac{CPPN}-only runs is smaller than when co-evolving the DSP graphs, as plotted in figure \ref{fig:eliteGenerationsAndGenomeSets} (\textit{right}).
The gradual coverage during the \textit{single-cell-win} runs can explain the lowest number of unique genomes shown in that figure.
Instead of distinguishing between individuals by their ID, where the differences could be only slight changes in e.g. connection weights, this plot is based on distinction between unique combinations of \ac{CPPN} and \ac{DSP} node and connection counts.

\subsection{Performance Against Pre-trained Reward Signals}\label{sec:Performance_Against_Pretrained_Reward_Signals}

\begin{figure}
    \centering
    \begin{tabular}{@{}c|@{}c}
        \includegraphics[width=0.49\textwidth]{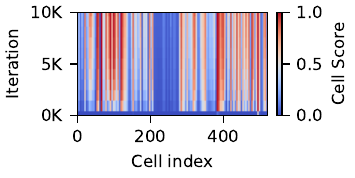}\label{cellScores_dur_05} &
        \includegraphics[width=0.49\textwidth]{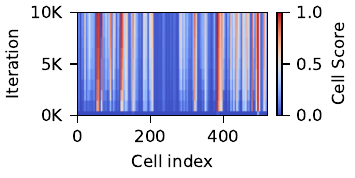}\label{cellScores_dur_CPPN-only_05}
    \end{tabular}
    \caption{Confidence scores declared by the \ac{YAMNet} \ac{DNN}, pre-trained on AudioSet classes (x-axis), averaged from the first 100 thousand iterations 
    of 10 runs. Results from a set of runs 
    where both \ac{CPPN} and \ac{DSP} genome counterparts are evolved can be seen 
    on the \textit{left}
    while 
    the \textit{right} map
    shows results from a set of runs 
    restricted to evolution of the \ac{CPPN} part of the genome, without evolving signal processing nodes.}
    \label{fig:cellScores_dur_05}
\end{figure}

Looking in more detail at the confidence levels assigned to each class, we see 
in figure \ref{fig:cellScores_dur_05}
that 
the \ac{YAMNet} classifier chosen in this iteration of our investigations assigned high scores to the sounds generated by our system across most classes.
There we can see again how the co-evolution of \acp{CPPN} with \ac{DSP} graphs achieves higher scores overall. The figure also reveals how the synthesiser struggles  in the range of classes between 214 and 276, which classify musical genres, rather than distinct sounds or instruments, such as “Pop music”, “Rhythm and blues”, “Flamenco”, etc. This is reasonable as the system is expected to generate sounds useful in the process of creating e.g. music, rather than entire musical compositions. Nonetheless it can be interesting to observe what the system came up with for those low-confidence classes, such as “Theme music”: 
a filter can be set in the online explorer (sec. \ref{sec:qualitative_evaluation}) to audition classes containing the phrase “music” while scrubbing through the runs with a slider.

\subsection{Genome Complexity}\label{sec:genome_complexity}
\begin{figure}[h!] 
    \centering
        \centering
        \includegraphics[width=0.6\textwidth]{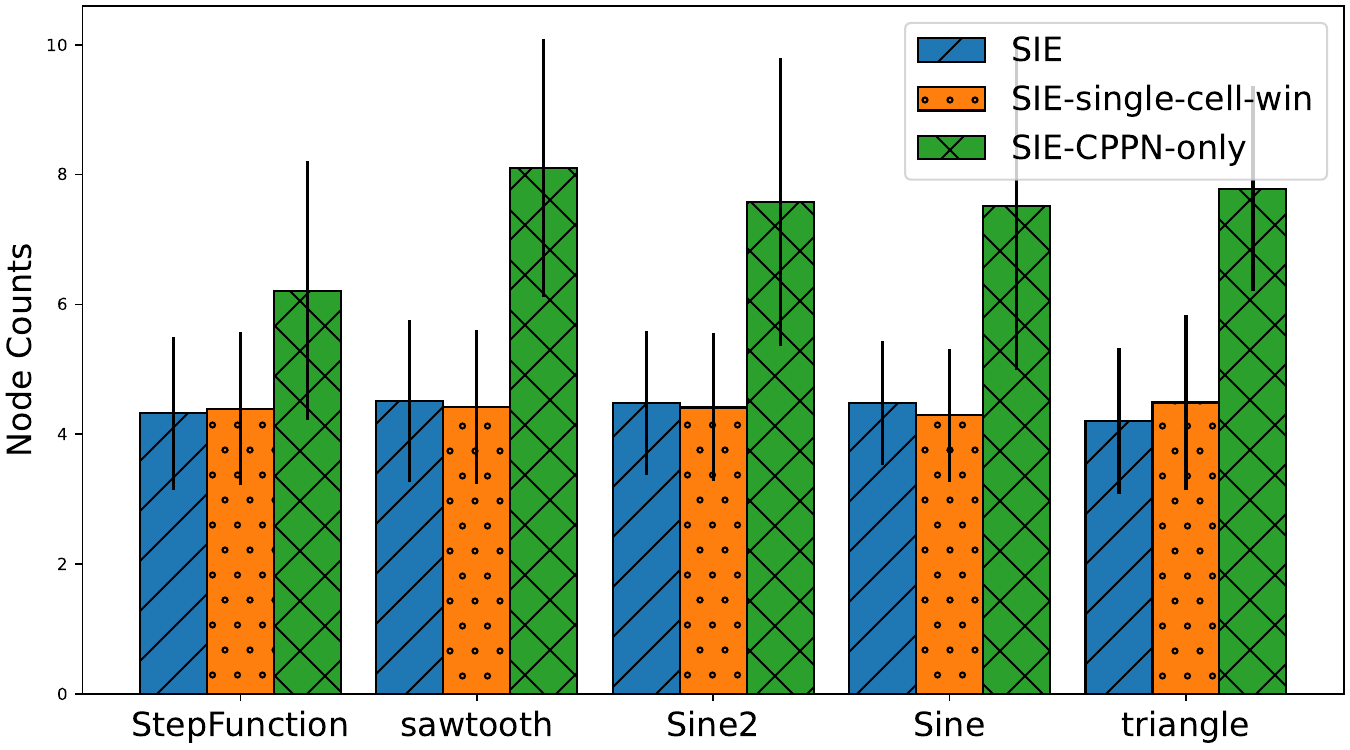}
        \label{genomeStatistics_node_type_count_CPPN}

    \bigskip
        \centering
        \includegraphics[width=0.95\textwidth]{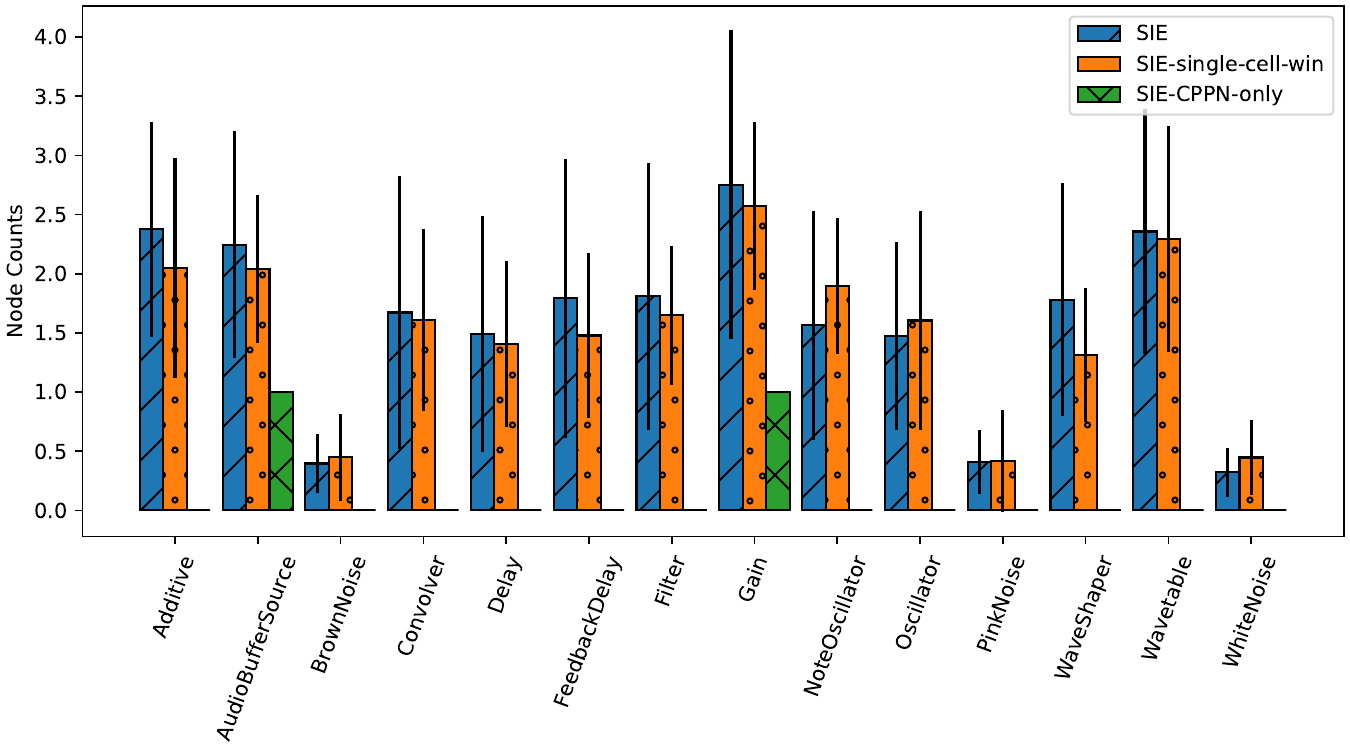}
        \label{genomeStatistics_node_type_count_asNEATPatch}
    \caption{Composition of \ac{CPPN} activation functions 
    (\textit{top})
    and \ac{DSP} graph node types 
    (\textit{bottom}),
    from the different evolution run variants. It can be observed in 
    the \ac{DSP} chart
    that the \ac{CPPN}-only variant does not evolve a \ac{DSP} graph and only includes a source node for receiving the pattern-signal from the single \ac{CPPN} output, and a gain node for passing it through to the output.}
    \label{fig:genomeStatistics_node_type_count}
\end{figure}

\begin{figure}
    \centering
    \begin{tabular}{@{}c|@{}c}
        \includegraphics[width=0.49\textwidth]{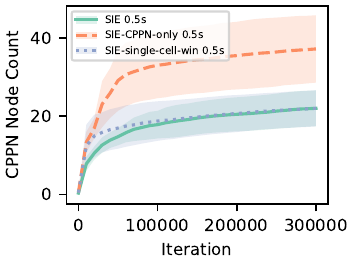}\label{genomeStatistics_CPPN_node_count} &
        \includegraphics[width=0.49\textwidth]{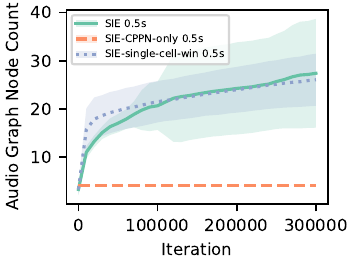}\label{genomeStatistics_audio_graph_node_count}
    \end{tabular}
    \caption{Genome complexity over the course of 10 \ac{QD} runs for each variant. \ac{CPPN} node counts are plotted 
    on the \textit{left}
    and \ac{DSP} graph node counts can be seen in 
    the plot on the \textit{right}
    (a flat line for the \ac{CPPN}-only variant 
    indicates that the 
    \ac{DSP} graph is not evolved.
    }
    \label{fig:genomeStatistics}
\end{figure}

The composition of audio graph nodes and \ac{CPPN} activation functions can be seen in figure \ref{fig:genomeStatistics_node_type_count}, where the prominence of the custom audiograph nodes (wavetable and additive synthesis, fig. \ref{fig:genomeStatistics_node_type_count}, \textit{bottom}) suggest that implementing other known techniques from the history of sound synthesis may be worthwhile. The distribution of \ac{CPPN} activation function types is quite uniform in all variants of our runs (fig \ref{fig:genomeStatistics_node_type_count}, \textit{top}).  
It’s also interesting to observe in 
the left plot of figure \ref{fig:genomeStatistics}
that the \ac{CPPN}-only runs resulted in more complex function compositions, likely to compensate for the lack of a co-evolving \ac{DSP} graph. This increased \ac{CPPN} complexity resulted in longer rendering times and thus increased durations of the evolution runs, as that part of the genome is more computationally expensive, with potentially many network activations required for each sample, as discussed in \cite{jonsson_system_2024}.

\subsubsection{Interplay between \ac{CPPN} and \ac{DSP} graphs}\label{sec:interplay_between_cppn_and_dsp_graphs}

Though (human or animal) brain’s processes are vastly more complex and less
linear than electronic synthesizer circuits, the notion of neuronal specialisation, from neuroscience and neurobiology, can serve as an analogy when considering the sound synthesis approach employed in our experiments, where \ac{CPPN} networks output wave patterns as signals for a \ac{DSP} graph, at different frequencies for different for different functions, such as for low frequency control signals and higher frequency sound sources. The human brain is organised into distinctive areas or regions, each of which is associated with specific functions. Those areas can operate at different frequencies, where a frequency from one region can modulate the frequency of another, through cross-frequency coupling (CFC). While also more complex than (modular) sound synthesis approaches, we can see some similarities with modulating sound signals by using control signals.

In our main experiments, the evolved genomes, which are rendered to sonic phenotypes, consist of a single \ac{CPPN} network and a corresponding \ac{DSP} graph, both evolved with \ac{NEAT}, where the \ac{CPPN} network can be used for multiple series of activations, one activation set for each frequency requested by the \ac{DSP} graph.  Over successive generations, \ac{NEAT} introduces mutations that can add new nodes and new connections. By slowly introducing complexity, \ac{NEAT} ensures a balance between exploring new structures and exploiting existing ones. 
This gradual complexification process allows for the evolution of intricate and sophisticated networks while avoiding the pitfalls of overly complex structures too soon in the evolutionary process. 

While both networks in our genome structure benefit from the gradual improvement enabled by NEAT, the question arose whether the \ac{CPPN}-based part was being tasked with too diverse tasks. Could it be beneficial to organise the wave-pattern-production into different structural regions---drawing inspiration from the brain analogy---allowing each to develop for more specific tasks? Instead of straining a single \ac{CPPN} against heterogeneous tasks, such as producing modulating control signals and audio rate signals (potentially to be modulated), separate \acp{CPPN} for each class of tasks could evolve more gradually, giving time for new innovations to mature and demonstrate their potential without being immediately outclassed by more general purpose networks, pressured to evolve more rapidly towards multiple tasks at the same time.

To test this hypothesis in the domain of sound synthesis,
we developed an extended version of the sound synthesis genome, where multiple \acp{CPPN} are instantiated, each for a specific range of frequencies requested by the corresponding \ac{DSP} graph: the ranges are defined such that frequencies below 20 Hz are rounded to the nearest integer and frequencies above (or equal to) 20 Hz are rounded to the nearest 10, so that the frequencies 9.1 Hz and 9.3 Hz are served by the same \ac{CPPN} and DSP inputs at 441 Hz and 444 Hz come from the same \ac{CPPN}.

To measure the effect of this modified genome structure, we conducted experiments composed of two sets of five \ac{QD} runs, for 9375 generations with a batch size of 32 each (9375 x 32 = 300K iterations), using the expanded behaviour space discussed in section \ref{sec:behaviour_space_expansion}. In the first set we used the original genome structure of a single \ac{CPPN} serving all frequencies requested by the co-evolving \ac{DSP} graph, and in the latter we used the genome type where one \ac{CPPN} network is instantiated for each frequency range. Comparing the results reported in table \ref{tab:genome-statistics-comparison}, we see that when using specialised \acp{CPPN}, each has on average only one third of the node count when compared to the usage of a single multi-purpose \ac{CPPN}, and the connection count is nearly half---while the number of requested frequency ranges by the \ac{DSP} is similar (10.18±1.66 vs. 12.37±3.34), and the usage of the 18 available \ac{CPPN} outputs is also close between those sets of runs (8.27±0.83 vs. 9.34±1.33). The \ac{DSP} network complexity is slightly higher when using multiple \acp{CPPN}, with higher variance (44.76±15.10 vs. 36.96±6.63), but the \ac{QD} score is also higher with those collections of much simpler \acp{CPPN} (1427.28±51.92 vs. 1408.58±73.10---out of a maximum 2605 in the expanded behaviour space: sec. \ref{sec:behaviour_space_expansion}).

This collection of genomes with simpler \ac{CPPN} networks, achieving comparable results to previous studies, 
can be better suited to further evolution, 
potentially within different measurement spaces, as they might tend to be more adaptable and robust to variations when offered new opportunities to evolve, such as in simulations of \ac{OEE}. This genome design can scale more gracefully as new frequency ranges or tasks are added. Instead of enlarging a single, increasingly complex \ac{CPPN}, new specialised \acp{CPPN} can be introduced, each optimised for new tasks.

\begin{table}[htbp]
\caption{Comparison of genome statistics between Single CPPN and One CPPN per frequency approaches}
\centering
\begin{tabular}{l|rr|rr}
\hline
& \multicolumn{2}{c|}{\textbf{Single CPPN}} & \multicolumn{2}{c}{\textbf{One CPPN per freq.}} \\
\textbf{Metric} & \textbf{Mean} & \textbf{Std Dev} & \textbf{Mean} & \textbf{Std Dev} \\
\hline
CPPNs & 1.00 & 0.00 & 15.80 & 5.23 \\
CPPN nodes & 32.68 & 4.10 & 10.30 & 2.81 \\
CPPN conns & 142.84 & 11.21 & 80.20 & 7.58 \\
Used outputs & 8.27 & 0.83 & 9.34 & 1.33 \\
Freqs. & 10.18 & 1.66 & 12.37 & 3.32 \\
DSP nodes & 36.96 & 6.63 & 44.76 & 15.10 \\
DSP conns & 64.05 & 12.18 & 79.48 & 27.07 \\
QD score & 1408.58 & 73.10 & 1427.28 & 51.92 \\
\hline
\end{tabular}

\label{tab:genome-statistics-comparison}
\end{table}

\subsection{Evolutionary Stepping-Stones}\label{sec_evolutionary_stepping_stones}
To assess how evolution leveraged the diversity promoted by our classifier, we conducted two measurements that explored the stepping stones across various classes.
One has been called \textit{goal switching} and defined as "the number of times during a run that a new class champion was the offspring of a champion of another class" in \cite{nguyen_innovation_2015,nguyen_understanding_2016}. From our runs we measured a mean of 21.7\textpm3.6 goal switches, 63.2\% of the 34.3\textpm4.5 mean new champions per class. This can be compared to the 17.9\% goal switches 
reported in \cite{nguyen_understanding_2016}.
Another way of measuring how the evolutionary paths flow though the stepping stones laid out by the classifier is to trace through the phylogenetic tree leading to each elite and then count how often its parent comes from a class different from the one it occupies. Counting from the current elites of each class at the end of the evolution runs, we found a mean of 44.9\textpm14.7 such occurrences.
In lieu of a visual phylogenetic presentation, the \textit{generation} slider of the evolution runs explorer (section \ref{sec:qualitative_evaluation}) can dynamically reveal how elites for each class come from different, often unrelated classes during the course of evolution.

\subsubsection{Context Switching}\label{sec:context_switching}

To further measure the ability of our QD experimental setup to model natural evolution through many, often surprising paths, where some organisms have unlikely ancestors \cite{stanley_why_2015}, we divided our classification into two broad groups: Musical and Non-musical.

The Musical group includes sounds that fall into the high level categories: \textit{Music-related human sounds (singing, rapping, etc.), Musical instruments (all types), Music genres and styles, Music techniques and performance styles, Music in various contexts (background music, soundtrack, etc.), Music production elements}.

Within the Non-musical context are sounds from classes that can be categorised within: \textit{Speech and human sounds (except singing and music-related vocalisations), Animal sounds, Environmental sounds (wind, water, etc.), Vehicle and mechanical sounds, Household and everyday sounds, Tools and machinery, Explosive and impact sounds, Material sounds (wood, glass, liquid, etc.), Miscellaneous sounds, Acoustic environments, Noise and sound qualities, Media devices (TV, radio)}.

Some sounds, like laughter or whistling, could be considered borderline cases. They are included in the musical category because they are often used rhythmically or melodically in musical contexts.

Results in table \ref{tab:goal-switches-statistics} show that while most goal switches occur within either context of the musical or non-musical group, 20\% of goal switches traverse the border between those broad groups (\textit{Context Switch/Dwell Ratio}). Tracing through the lineage of each current elite shows that 14\% of parents in the ancestry come from across that border (\textit{Cross Border Parent Ratio}). Those are signs of the lineages traversing unlikely stepping stones on their paths to the current elites.

\begin{table}[htbp]
\caption{Goal and Context Switch Statistics}
\centering
\begin{tabular}{lrr}
\hline
\textbf{Metric} & \textbf{Mean} & \textbf{Std Dev} \\
\hline
Average Champion Count & 29.11 & 1.41 \\
\hline
Average Goal Switch Count & 24.61 & 1.64 \\
Average Context Switch Count & 4.50 & 0.53 \\
Average Context Dwell Count & 22.93 & 1.02 \\
Context Switch/Dwell Ratio & 0.20 & 0.02 \\
\hline
Parents from Same Group & 73.90 & 8.23 \\
Parents from Another Group & 10.27 & 2.61 \\
Cross Border Parent Ratio & 0.14 & 0.05 \\
\hline
\end{tabular}
\label{tab:goal-switches-statistics}
\end{table}

\begin{figure}
\centering
\includegraphics[width=\textwidth]{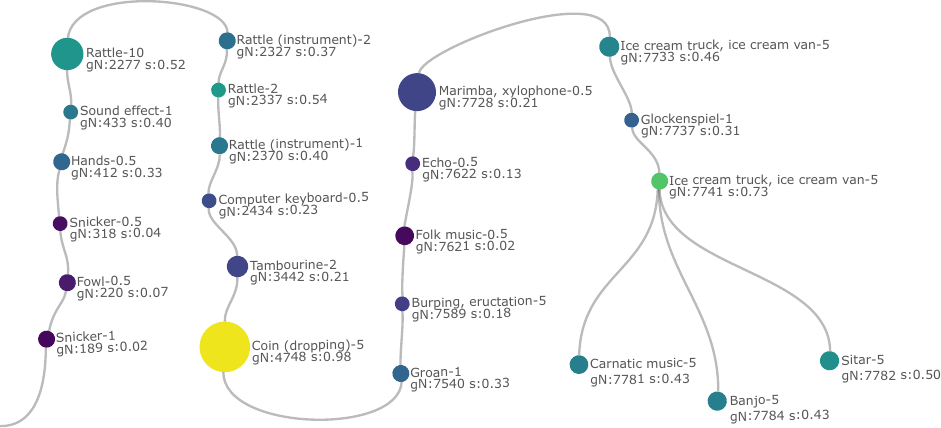}
\caption{Stepping stones across contexts in one lineage towards the musical classes Carnatic music, Banjo and Sitar. The label "Echo-0.5 gN:7622 s:0.13" signifies a sound from the YAMNet class Echo, for the duration of half a second, which became an elite at generation 7622, by receiving the confidence score 0.13 (out of a maximum 1).}
\label{fig:lineage}
\end{figure}

Figure \ref{fig:lineage} shows context switches in a lineage towards elites in three classes categorised in the musical group. The number of context switches in that particular lineage is double the average of \textit{Parents from Another Group} (10.27±2.61) in table \ref{tab:goal-switches-statistics}.
Stepping stones through classes within each group are concealed in figure \ref{fig:lineage}, to emphasise the cross-context traversals---according to \textit{Parents from Same Group} in table \ref{tab:goal-switches-statistics}, they are on average seven times more frequent (73.90±8.23).
Dynamic renditions of lineages from the \ac{QD} runs discussed here can be explored interactively and elite sounds throughout auditioned at a web page accompanying this article\footnote{
    Web page accompanying this article:\newline\url{https://www.uio.no/ritmo/english/people/phd-fellows/bthj/publications/quality-diversity-search-in-sound-generation.html}
}.

\subsection{Abandoning Diversity}
Growth of genome complexity seems to have stayed within reasonable limits, even when \acp{CPPN} were left alone to the task of performing against the classifier (fig. \ref{fig:genomeStatistics}).
An exception to this is when we experimented with abandoning diversity and adopting single objectives. 
Though the benefit of diversity has been demonstrated \cite{nguyen_innovation_2015,nguyen_understanding_2016}, we investigated how a similar experiment fares in the sound domain. To that end, we selected 10 classes\footnote{
    Single-class runs were performed on the classes Aircraft, Banjo, Beatboxing, Boom, Choir, Dubstep, Fusillade, Mandolin, Synthetic singing and Whistling.
} 
as single objectives of separate runs 
and compared the performance and genome complexity with the performance from the \ac{QD} runs on those same classes. 

\begin{figure}
    \centering
    \begin{tabular}{@{\hspace{0.01\textwidth}}c@{\hspace{0.01\textwidth}}c@{}c}
        \includegraphics[width=0.33\textwidth]{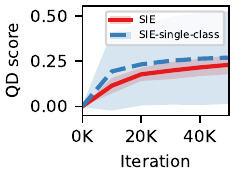}\label{fig:qdScores-single_vs_multiclass} &
        \includegraphics[width=0.33\textwidth]{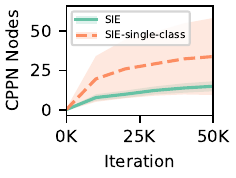}\label{genomeStatistics_CPPN_node_count-single_vs_multiclass} &
        \includegraphics[width=0.33\textwidth]{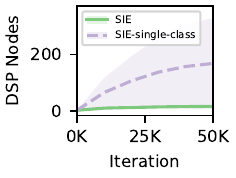}\label{genomeStatistics_audio_graph_node_count-single_vs_multiclass}
    \end{tabular}
    \caption{
    The \textit{first} plot shows
    performance scores from single-objective vs. multi-objective \ac{QD} runs, averaged from a set of randomly selected classes.
    The \textit{second} and \textit{third} plots show how
    genome complexity 
    developes during 
    single- and multi-objective \ac{QD} runs, 
    in terms of 
    \ac{CPPN} and \ac{DSP} graph node counts.
    }
    \label{fig:genomeStatistics-single_vs_multiclass}
\end{figure}

Interestingly, although the performance in single objective runs is higher on average than in multi-class runs, as shown in 
the \textit{first} plot in figure \ref{fig:genomeStatistics-single_vs_multiclass},
the difference is accompanied by a higher level of deviation and much higher genome complexity. 
The \textit{second} and \textit{third} plots in figure \ref{fig:genomeStatistics-single_vs_multiclass}
indicate that the CPPN and DSP node counts in genomes from single objective runs are significantly higher than those of genomes from the same set of classes in \ac{QD} runs.
The computational effort required for the complex genomes evolved during the single class runs limited our iteration count to 50 thousand, 1/6th of the iterations performed for the 
baseline \ac{QD} runs. The unexpected result of higher performance from the single-class runs may be attributed to the narrow set of chosen classes; this experiment could benefit from further investigation.






\subsection{Temporal Pattern Revelation and Classifier Characteristics}
Although half a second sounds were the most prevalent renditions of successful individuals in our manual parameters search (section \ref{sec_approach}), comparing sets of runs with two large variations in the duration of the evaluated phenotypes was interesting. We chose to compare runs evaluating half a second renditions of the evolved genomes with a set of runs evaluating ten-second renditions. One motivation for the choice of the longer duration, is that "\ac{YAMNet} is trained on 1,574,587 10-second YouTube soundtrack excerpts from within ... AudioSet"\footnote{\ac{YAMNet} release announcement:\newline \url{https://groups.google.com/g/audioset-users/c/U71MxTdHqkU}}. 
While \ac{CPPN}-only runs achieved less overall confidence when rendering 0.5s sounds for evaluation by the classifier, as can be seen
on the \textit{left} of figure \ref{fig:qdScores},
we hypothesised that allowing the classifier to sample in more detail the patterns developed by the \acp{CPPN}, when processing more frames over a longer duration, would result in higher confidence. The opposite turned out to be the case, where \ac{CPPN}-only runs, rendering 10s sounds for evaluation achieved a lower QD score than corresponding runs rendering 0.5s sounds. Perhaps the lack of \ac{DSP} becomes more significant in the evaluation of longer duration sounds. Duration has little effect when \ac{DSP} graphs evolve alongside the \acp{CPPN}, as 
the \textit{right} plot in figure \ref{fig:qdScores}
shows.

\subsubsection{Expanding Behavior Space into the Temporal Dimension}\label{sec:behaviour_space_expansion}
The sound synthesis engine we base our experiments on can render sounds from the underlying genome at any duration and resolution, by increasing the number and resolution of the wave-generating \ac{CPPN} networks. Surprisingly, we observed that genomes declared as elites for one class and duration, such as 0.5s in our main experiments (section \ref{sec_parameter_search}), often did not render convincing results in that class context at other durations. To investigate further that distinctive aspect of sound, the temporal dimension, we configured experiments where several sound durations were added as a dimension to the behaviour space, perpendicular to the classification dimension: so for each class there are opportunities to become an elite when rendered at 0.5, 1, 2, 5 and 10 seconds.

\begin{figure}
\centering
\includegraphics[width=\textwidth]{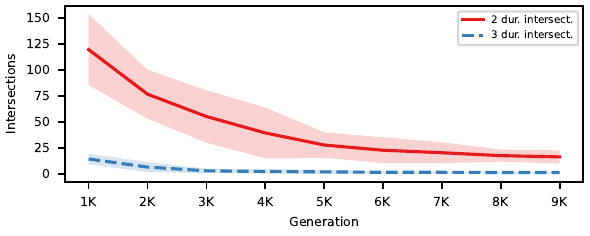}
\caption{Genome intersections through temporal variations; counting the occurrence of genomes responsible for elites at two or three different sound durations}
\label{fig:duration-intersections}
\end{figure}

Intuitively, one might assume that a genome rendering an elite for the class Violin at the duration of 1 second, should also render a sound at 5 seconds that would maintain an elite status for that class. Measuring how often a single genome produces phenotypes that are elites at different durations, reveals that this is seldom the case. Referring to table \ref{tab:genome-sets-across-durations}, we can see that while each duration-cell along the temporal dimension contains more than 300 unique genomes after one thousand generations, figure \ref{fig:duration-intersections} reveals that only around 120 genomes are responsible for elite sounds at two different durations, and only a handful at three durations. The plot in that figure shows how those duration intersections diminish as the evolution progresses, even though the table shows an increase in the number of unique genomes at each duration, indicating specialisation at the niches defined by duration and how the revelation of nuances in the patterns produced by the \acp{CPPN} seem to present new opportunities for elite performance against the classifier. Elite sounds for each class, at different durations, are more often than not audibly distinct, so those are not just slight variations.

A mean of 29.1\textpm1.4 goal switches was measured from five runs at this configuration (statistically significant from the 21.7\textpm3.6 goal switches, reported in section \ref{sec_evolutionary_stepping_stones} (according to a Poisson rate test,
p $<$ 0.001).
Counting the number of parents from different classes throughout the phylogenetic tree results in a mean of 74.0\textpm5.0, nearly double that of 44.9\textpm14.7 reported in \ref{sec_evolutionary_stepping_stones} of such occurrences, which can be attributed to a higher number of cells in the behavious space (521 cells vs. 521\texttimes5 (duration cells) = 2605 cells in total).
The difference in goal switches being far from double, though significant, can be interpreted as confirming the observed path of specialisation within the vaster behaviour space, only by adding the dimension of temporal sound variation.

\begin{table}[h]
\caption{Count of unique genomes within each temporal dimension, which has been added orthogonally to each class, across the generations of evolutionary runs}
\centering
\begin{tabular}{r|cc|cc|cc|cc|cc|cc}
\hline
\multirow{2}{*}{Generation} & \multicolumn{2}{c|}{0.5s} & \multicolumn{2}{c|}{1s} & \multicolumn{2}{c|}{2s} & \multicolumn{2}{c|}{5s} & \multicolumn{2}{c|}{10s} \\
 & Mean & SD & Mean & SD & Mean & SD & Mean & SD & Mean & SD \\
\hline
1K & 369.4 & 10.50 & 384.8 & 16.07 & 375.8 & 15.22 & 364.8 & 25.17 & 360.6 & 17.51 \\
2K & 398.2 & 7.55 & 406.6 & 14.05 & 398.6 & 17.17 & 391.6 & 23.52 & 380.2 & 22.51 \\
3K & 410.4 & 6.56 & 416.8 & 9.77 & 416.6 & 12.89 & 410.2 & 14.88 & 396.6 & 21.24 \\
4K & 419.2 & 2.79 & 420.6 & 5.85 & 419.2 & 14.01 & 420.4 & 16.62 & 410.0 & 14.86 \\
5K & 422.4 & 4.45 & 423.4 & 11.36 & 425.6 & 14.57 & 422.2 & 15.61 & 412.0 & 17.29 \\
6K & 428.4 & 5.16 & 429.4 & 10.07 & 424.8 & 17.52 & 428.4 & 17.23 & 421.4 & 12.99 \\
7K & 427.8 & 4.35 & 434.6 & 10.93 & 429.6 & 15.45 & 435.2 & 14.29 & 428.8 & 11.50 \\
8K & 426.6 & 2.65 & 434.2 & 10.15 & 432.6 & 14.22 & 439.8 & 13.83 & 432.8 & 10.24 \\
9K & 430.4 & 3.38 & 436.2 & 10.70 & 433.2 & 12.48 & 442.6 & 13.94 & 436.6 & 10.15 \\
\hline
\end{tabular}
\label{tab:genome-sets-across-durations}
\end{table}

\subsection{Access to Sound Objects and their Application}
\label{sec:qualitative_evaluation}

\begin{figure}
    \centering
    \begin{tabular}{c|c}
        \includegraphics[width=0.238\textwidth]{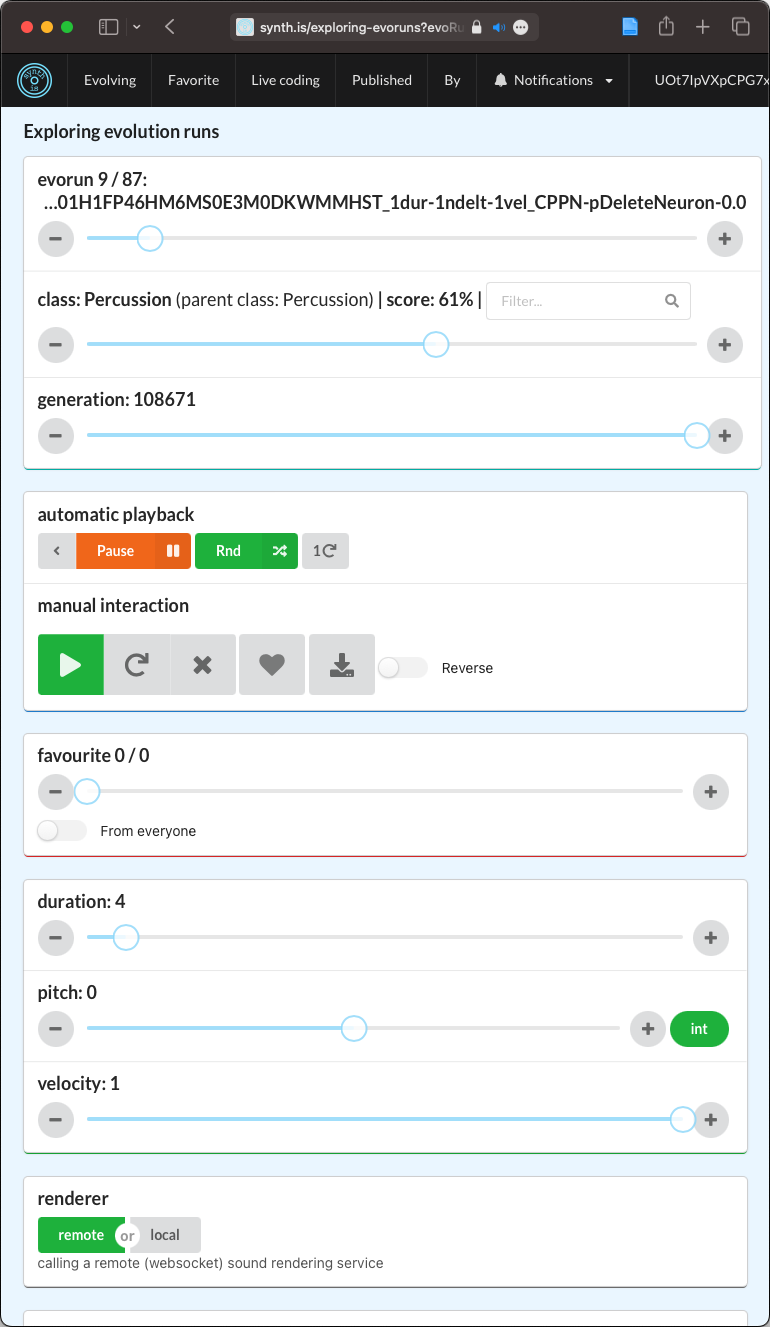}\label{fig:evoruns_explorer} &
        \includegraphics[width=0.73\textwidth]{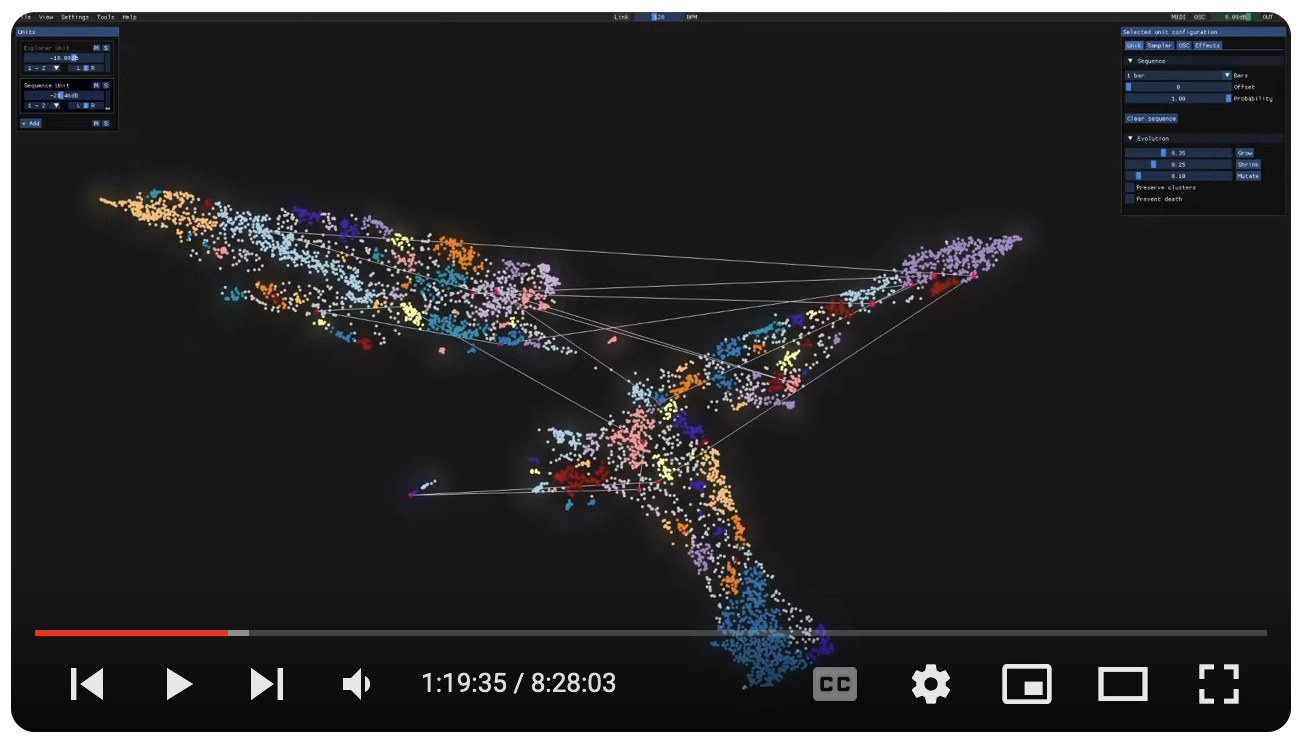}\label{fig:FNAS}
    \end{tabular}
    \caption{(\ref{fig:evoruns_explorer}) Evolution runs explorer, where it is possible to scrub through evolution runs, their classes and generations. The sound properties duration, pitch and velocity can be changed and favourites can be collected.
    (\ref{fig:FNAS}) Live streams (recorded) of automated, evolutionary sequences through sounds rendered from the evolutionary runs discussed in this paper, as one way of experiencing and qualitatively evaluating the generated artefacts. The sequencing is peformed by the experimental sampler AudioStellar.}
    \label{fig:qualitativeDissemination}
\end{figure}

We have facilitated open access to the generated artifacts through different means.
Those include an evolution runs explorer\footnotemark{}, 
depicted in figure \ref{fig:evoruns_explorer}. Final elites from all runs have also been rendered to  
WAV
files, which have been included in the accompanying datasets \cite{jonsson_supporting_2024,jonsson_extended_2024}. The sound 
objects 
in the pre-rendered files reflect the render-settings used 
to evaluate the corresponding genome that became an elite.
The online explorer\footnotemark[\value{footnote}] provides greater flexibility as it dynamically renders sounds with the default settings, but the interface also enables users to modify these settings. This modification can potentially reveal other intriguing sonic behaviors from the same genome.
\footnotetext{Evolution runs explorer: \url{https://synth.is/exploring-evoruns}}

As part of our investigation into the applicability of the discovered 
artefacts 
for creating other art, we loaded subsets of them into the experimental sampler AudioStellar \cite{garber_audiostellar_2021} and used that software to drive 
evolutionary sequences through the phenotypes. 
A playlist of live-stream recordings showcasing evolutionary sequences using sounds discovered by \ac{QD} runs is accessible 
online\footnote{
    Playlist with evolutionary sequences through sounds discovered by \ac{QD} runs:\newline\url{https://youtube.com/playlist?list=PLSYAaR-xYhEXk0czfHYKJSWmZ8vG35xEN}
}.
These compositions are largely automated, with human input limited to initial settings like evolutionary sequencing rates and fundamental sound effects. Nonetheless, they demonstrate the potential of the discovered sound objects to inspire creative endeavors. It is thought-provoking to consider if a human, given the same dataset, could craft more aesthetically pleasing arrangements with these sonic artefacts. 
We encourage the reader to obtain a copy of the files and engage in such experimentation \cite{jonsson_supporting_2024,jonsson_extended_2024}.

\section{Conclusion and Future Work}

Applying the combination of a diversity-promoting algorithm with selection pressure from a classifier reward signal to the search for 
sounds has been demonstrated to be a viable approach by the results discussed in this paper. Furthermore, the distinct approach to sound synthesis employed in this work has achieved high 
confidence from  a \ac{DNN} based classifier 
in most classes. 
High-scoring sounds are, in many cases, not the most realistic representatives of their class, especially when considering non-musical instrument classes, which can be attributed to how \ac{DNN}s are easily fooled \cite{nguyen_deep_2015}. 
Other recently proposed classification approaches may be more robust and could be worth investigating \cite{gong_ssast_2022,huang_masked_2022}, but classification robustness may not be the most sought-after quality in a creative system.
With a focus on the diversity-promoting attribute of the selection pressure applied in this investigation,
the diverse and innovative sound objects generated suggest that further explorations may be based on this system.
The intent would be to broaden the range of potential discoveries within the sonic domain.

Adopting 
\ac{YAMNet} as a 
classifier for sounds, to provide selection pressure for a \ac{QD} algorithm, was a 
step towards investigating 
a simple version of the Innovation Engine algorithm in the domain of sounds. Further explorations may include expanding the behaviour space to search beyond predefined classes. This can be done by combining the feature extraction ability of a \ac{DNN}, such as the one employed in this work, with \ac{DR}, as has been done in the visual domain in \cite{mccormack_quality-diversity_2022}. Extracting features with \acp{VAE} \cite{kingma_auto-encoding_2014}, and applying a clustering algorithm in the resulting latent space to define niches, as stepping stones during \ac{QD} search, is another approach \cite{mccormack_creative_2023} worth exploring further in the domain of sounds. While \acp{VAE} require a training set, limiting the 
behaviour space to explore, periodically retraining a 
\ac{DR} algorithm 
on discovered 
sound objects 
could enable 
autonomous and unsupervised discovery of the space of sounds which the 
generative 
system is able to render, without prior training, as proposed in \cite{cully_autonomous_2019,grillotti_unsupervised_2022}. Human intuition can also be leveraged to derive semantically meaningful diversity in the search space, as studied in \cite{ding_quality_2023}, which can be especially important  when generating sonic material leading 
to interesting discoveries according to individual aesthetics.

In the broadest sense, the concept of instruments has evolved from being a mere means to an end to a starting point for 
a journey into the unknown \cite[p. 49]{magnusson_sonic_2019}. 
The evolutionary system explored here is not intended as an instrument for serving requests from preconceived ideas but rather as a tool for discovering 
interesting sound objects that can steer the creative journey.
The sound artefacts generated by our system, as discussed in this paper,
are intended to facilitate or inspire the creation of further sonic art. This is different from the visual artefacts produced by many generative systems, which are often seen as standalone pieces without further utility.
Instead of a top-down approach---where the end goals and characteristics of the desired sound are pre-defined---our method encourages a bottom-up process of exploration. This reflects the evolutionary path of human development, where cognitive skills have been shaped by the very tools that humans have uncovered. This echoes the saying, “the tool writes the toolmaker as much as the toolmaker writes the tool” (\cite{davis_replications_1996} as cited in \cite[p. 5]{magnusson_sonic_2019}). 
An instrument that promotes such exploratory discovery can enable us to continue on our path of evolution by developing human abilities through technology. 

\backmatter





\bmhead{Acknowledgements}
This work was performed on Educloud Fox – High Performance Computing cluster, owned by the \href{https://www.usit.uio.no/english/index.html}{University of Oslo IT Department}.
Additional computations (for the \hyperref[sec:qualitative_evaluation]{evolution runs explorer}) were performed on the \href{https://www.nrec.no/}{Norwegian Research and Education Cloud (NREC)}, using resources provided by the University of Bergen and the University of Oslo.
Data storage and analysis was performed on resources provided by \href{https://www.sigma2.no/}{Sigma2} - the National Infrastructure for High-Performance Computing and
Data Storage in Norway, project NS9648K.
This work was supported by the Research Council of Norway through its Centres of Excellence scheme, project number 262762.
We would like to thank Bjarni Þór Jónsson for suggesting the application of Git for preserving fine-grained historical data from our simulations.



\section*{Declarations}

\subsection*{Funding}
This work was supported by the Research Council of Norway through its Centres of Excellence scheme, project number 262762.

\subsection*{Conflict of interest}
On behalf of all authors, the corresponding author states that there is no conflict of interest.

\subsection*{Ethics approval and consent to participate}
Not applicable.

\subsection*{Consent for publication}
Not applicable.

\subsection*{Data availability}
Data from evolutionary runs and analysis results are available in public repositories: \cite{jonsson_supporting_2024, jonsson_extended_2024}.

\subsection*{Materials availability}
Not applicable.

\subsection*{Code availability}
Code for replicating the simulations and analysis discussed in this article is available in public repositories: \cite{jonsson_synth-iskromosynth_2024,jonsson_synth-iskromosynth-cli_2024,jonsson_synth-iskromosynth-evaluate_2024,jonsson_synth-iskromosynth-render_2024}.

\subsection*{Author contribution}
All authors contributed to the study conception and design. 
Simulation design, execution, analysis and corresponding software development were performed by Björn Þór Jónsson. The first draft of the manuscript was written by Björn Þór Jónsson and all authors commented on previous versions of the manuscript.

\bibliography{ref-zotero-BibTex,kyrre}

\begin{acronym}[CPPN] %

\acro{EA}{Evolutionary Algorithm}
\acrodefplural{EA}[EAs]{Evolutionary Algorithms}

\acro{IEC}{Interactive Evolutionary Computation}

\acro{CPPN}{Compositional Pattern Producing Network}
\acrodefplural{CPPN}[CPPNs]{Compositional Pattern Producing Networks}

\acro{CSSN}{Compositional Sound Synthesis Network}
\acrodefplural{CSSN}[CSSNs]{Compositional Sound Synthesis Networks}

\acro{NEAT}{NeuroEvolution of Augmenting Topologies}
\acro{DSP}{Digital Signal Processing}

\acro{VAE}{Variational Auto Encoder}
\acrodefplural{VAE}[VAEs]{Variational Auto Encoders}

\acro{DNN}{Deep Neural Network}
\acrodefplural{DNN}[DNNs]{Deep Neural Network}

\acro{QD}{Quality Diversity}
\acro{MAP-Elites}{Multi-dimensional Archive of Phenotypic Elites }
\acro{YAMNet}{Yet Another Mobile Network}
\acro{HPC}{High Performance Computing}
\acro{gRPC}{gRPC Remote Procedure Calls}
\acro{WAV}{Waveform Audio File Format}

\acro{DR}{dimensionality reduction}

\acro{PCA}{Principal Component Analysis}
\acro{UMAP}{Uniform Manifold Approximation and Projection}
\acro{OEE}{Open-Ended Evolution}
\acro{OE}{Open-Endedness}
\acro{CC}{Computational Creativity}

\end{acronym}

\end{document}